\documentclass[prx,onecolumn,preprintnumbers,superscriptaddress,11pt]{revtex4} 

\usepackage{graphicx}
\usepackage{bm}
\usepackage{amsmath}
\usepackage{amssymb}
\usepackage{amsfonts}
\usepackage{fancyhdr}
\usepackage{slashed}
\usepackage{latexsym,epsfig,bbm}

\newcommand{\bra}[1]{\langle{#1}|}
\newcommand{\ket}[1]{|{#1}\rangle}

\newcommand{\braket}[2]{\langle{#1}|{#2}\rangle}

\newcommand{\figref}[1]{Fig.~\ref{#1}}

\usepackage{color}
\definecolor{blue}{rgb}{0,0.2,1}

\definecolor{red}{rgb}{0.9,0,0}

\newcommand{\past}[1]{\overleftarrow{#1}}
\newcommand{\fut}[1]{\overrightarrow{#1}}
\newcommand{\tpast}{t_{\overleftarrow{0}}}
\newcommand{\Tpast}{T_{\overleftarrow{0}}}
\newcommand{\tfut}{t_{\overrightarrow{0}}}
\newcommand{\Tfut}{T_{\overrightarrow{0}}}

\begin{document}

\title{Memory-efficient tracking of complex temporal\\ and symbolic dynamics with quantum simulators}

\author{Thomas J. Elliott}
\email{physics@tjelliott.net}
\affiliation{Complexity Institute, Nanyang Technological University, Singapore 637335}
\affiliation{School of Physical and Mathematical Sciences, Nanyang Technological University, Singapore 637371}

\author{Andrew J. P. Garner}
\affiliation{Institute for Quantum Optics and Quantum Information, Austrian Academy of Sciences, Boltzmanngasse 3, A-1090 Vienna, Austria}
\affiliation{Centre for Quantum Technologies, National University of Singapore, 3 Science Drive 2, Singapore 117543}

\author{Mile Gu}
\email{mgu@quantumcomplexity.org}
\affiliation{School of Physical and Mathematical Sciences, Nanyang Technological University, Singapore 637371}
\affiliation{Complexity Institute, Nanyang Technological University, Singapore 637335}
\affiliation{Centre for Quantum Technologies, National University of Singapore, 3 Science Drive 2, Singapore 117543}

\date{\today}

\begin{abstract}
Tracking the behaviour of stochastic systems is a crucial task in the statistical sciences. It has recently been shown that quantum models can faithfully simulate such processes whilst retaining less information about the past behaviour of the system than the optimal classical models. We extend these results to general temporal and symbolic dynamics. Our systematic protocol for quantum model construction relies only on an elementary description of the dynamics of the process. This circumvents restrictions on corresponding classical construction protocols, and allows for a broader range of processes to be modelled efficiently. We illustrate our method with an example exhibiting an apparent unbounded memory advantage of the quantum model compared to its optimal classical counterpart.
\end{abstract}
\maketitle 

\section{Introduction}

Continuous-time stochastic processes are omniprescent across the sciences. They are used to model a rich and diverse range of systems~\cite{yu2010hidden}, such as speech recognition~\cite{levinson1986continuously, rabiner1989tutorial}, financial time-series~\cite{bulla2006stylized}, neuronal spike trains~\cite{gerstner2002spiking, marzen2015time}, gene recognition~\cite{kulp1996generalized}, Internet traffic~\cite{yu2002hidden}, and geophysical processes~\cite{garavaglia2011earthquake}. Given this broad applicability, our ability to study, simulate, and make predictions using such models is of great import. However, simulations of such models can become highly resource-intensive, in part due to their continuous nature. The information that must be tracked about the past of the process typically diverges with increased precision~\cite{marzen2015time, marzen2015informational, marzen2017informational}. 

\emph{Computational mechanics}~\cite{shalizi2001computational, crutchfield2012between} provides a toolset that may be employed in optimising the use of certain resources. Stemming from notions of structural complexity in stochastic processes~\cite{grassberger1986toward, zambella1988complexity, crutchfield1989inferring}, it prescribes a framework for obtaining minimal memory predictive models of a process. This has been applied to a panoply of discrete-time processes~\cite{crutchfield1997statistical, tino1999extracting, palmer2000complexity, clarke2003application, park2007complexity, li2008multiscale, haslinger2010computational, kelly2012new, marzen2015time}, but only recently have similar studies been made for continuous-time processes, in restricted settings~\cite{marzen2017informational, marzen2017structure}.

In parallel, the field of \emph{quantum computational \mbox{mechanics}} has emerged~\cite{gu2012quantum, mahoney2016occam, riechers2016minimized, aghamohammadi2016ambiguity, palsson2017experimentally, suen2017classical, garner2017provably, aghamohammadi2017extreme, elliott2018superior, binder2018practical, ghafari2017observing, thompson2018causal, aghamohammadi2018extreme}. A central result arising from these works is that quantum models of stochastic processes can operate whilst tracking less information about the past than even the optimal classical models~\cite{gu2012quantum}. This quantum advantage can be immense, and the gap between quantum and optimal classical memory requirements can grow unbounded~\cite{garner2017provably, aghamohammadi2017extreme, elliott2018superior}. As with classical computational mechanics, the focus has largely been on discrete-time processes, and hitherto the quantum computational mechanics of continuous-time processes has been restricted to tracking only a limited set of temporal dynamics (renewal processes), where the times between consecutive emissions are all drawn from the same distribution~\cite{elliott2018superior}. Nevertheless, it was found that quantum models of such processes can exhibit unbounded advantages, requiring only finite memory to predict processes that classically need infinite past information to model.

Here, we develop these concepts into a systematic prescription for producing quantum models that simultaneously track temporal and symbolic dynamics  in complex stochastic processes, for which the information about the past that must be stored for future prediction is compressed beyond classical limits. After introducing the computational mechanics framework, we provide a protocol that constructs quantum models of multi-symbol, multi-state stochastic processes in both the discrete- and continuous-time regimes, for which causal equivalence relations are automatically satisfied. This represents the most general construction for quantum models of complex stochastic processes thus far, going far beyond existing quantum constructions and opening the possibility to study a much broader range of complex systems. Moreover, as our construction protocol can be applied to general temporal and symbolic dynamics, it also covers a broader range of systems than existing construction protocols for optimal classical models of stochastic processes. Further, our models exhibit an entropic quantum advantage in the memory required over optimal classical models, even in such cases where systematic construction protocols for these optimal classical models are unknown. We illustrate these features with an example showcasing these advantages, and conclude by highlighting avenues for future development.

\section{Framework}

We consider continuous-time, discrete event stochastic processes. Such processes are characterised by a sequence $(x_{\bm{n}},t_{\bm{n}})$ detailing what is observed, and when. The emitted symbols $x_n\in\mathcal{A}$ denote the events, while $t_n$ records the time elapsed between events $n-1$ and $n$. Sequences occur with probabilities drawn from $P(X_{\bm{n}},T_{\bm{n}})$~\cite{khintchine1934korrelationstheorie} (upper-case denotes random variables, and lower-case their corresponding realisations). We use shorthand notation ${\bm x}_n=(x_n,t_n)$, and denote contiguous strings of observations by the concatenation ${\bm x}_{l:m}={\bm x}_l{\bm x}_{l+1}\ldots{\bm x}_{m-1}$. We restrict our attention to stationary processes, wherein $P({\bm X}_{0:L})=P({\bm X}_{s:s+L})\forall s,L\in\mathbb{Z}$. This framework accommodates emissions that take place either as instantaneous events separated by times $t_n$, or as continuous emissions with dwell time $t_n$. We focus primarily on the former, and later discuss how our results may be modified for the latter. Further, though our primary focus will be on continuous-time processes, many of the results can be applied to temporal discretisations of such processes, and we will provide an explicit construction for these coarse-grained analogues.

The observation sequence can be partitioned into past and future. We take 0 as the current emission step, such that $x_0$ is the next symbol to be emitted, and define $\tpast$ ($\tfut$) as the time since the last (until the next) emission, such that $t_0=\tpast+\tfut$. We delineate the past as $\past{\bm{x}}={\bm x}_{-\infty:0}(\emptyset,\tpast)$ ($\emptyset$ signifies the event $x_0$ is currently unknown), and the future as  $\fut{\bm{x}}=(x_0,\tfut){\bm x}_{1:\infty}$~\cite{marzen2017structure}.

\begin{figure}
\includegraphics[width=\linewidth]{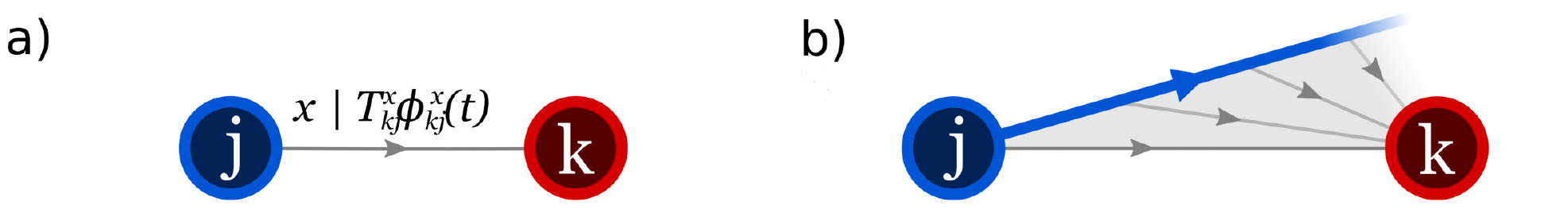}
\caption{{\bf Diagrammatic representation of continuous-time processes.} Continuous-time, discrete event stochastic processes may be drawn as edge-emitting hidden semi-Markov models, where (a) the system transitions between a set of hidden modes while emitting symbols $x$, with dynamics defined by $T_{kj}^x\phi_{kj}^x(t)$. (b) The temporal dynamics can be tracked by their unpacking from the modes into hidden causal states (thick line) that store relevant information about the mode and time since last emission.}
\label{figdiagram}
\end{figure}

Such processes can be represented by edge-emitting hidden semi-Markov models (eeHSMM)~\cite{marzen2017structure}. These are defined by a set of hidden modes $\{g\}$, an emission alphabet $\mathcal{A}$, and a transition dynamic $T_{kj}^{x}\phi_{kj}^{x}(t)$. The transition dynamic describes the probability density that the system, upon transitioning to mode $j$, will subsequently reside in this mode for a time $t$, at which point it will transition to mode $k$ while emitting symbol $x$. The $\phi_{kj}^x(t)$ are normalised, such that $T_{kj}^x$ describes the total probability that the system transitions from $j$ to $k$ while emitting $x$ without reference to the time. We can represent such models diagrammatically [\figref{figdiagram}(a)]. Semi-Markov~\cite{yu2010hidden} refers to the property that the transition dynamic depends only on the current mode and dwell time, such that the causal pair $(g,\tpast)$ gives the fullest possible description for predicting the future of the process that may be obtained from past observations.

We desire models of a process that can faithfully reproduce future statistics given a particular past, and that are \emph{causal}, i.e.~contain no information about the future that is not obtainable from the past~\cite{thompson2018causal}. Computational mechanics~\cite{shalizi2001computational, crutchfield2012between} provides a pathway for determining the optimal (storing minimal information) classical models. The basis of computational mechanics are causal states $\mathcal{S}$, equivalence classes agglomerating pasts with identical future predictions~\cite{crutchfield1989inferring}. Two pasts $\past{\bm{x}}$ and $\past{\bm{x}}'$ belong to the same causal state (are causally equivalent $\sim_e$) iff they have identical conditional future probabilities:

\begin{equation}
\past{\bm{x}}\sim_e\past{\bm{x}}'\Leftrightarrow P(\fut{\bm{X}}|\past{\bm{x}})=P(\fut{\bm{X}}|\past{\bm{x}}').
\end{equation}

It has been proven that using these causal states as the hidden states of a model provides the optimal classical predictive representation~\cite{shalizi2001computational}. For discrete-time processes, these form edge-emitting hidden Markov models, while for continuous-time processes one has a continuum of hidden states the system traverses along, jumping into a hidden `start' state (mode of an eeHSMM) upon emission [\figref{figdiagram}(b)]. The optimal classical predictive models are called $\varepsilon$-machines, and are unifilar: given knowledge of a prior causal state and the observation sequence since, the present causal state is known with certainty~\cite{shalizi2001computational}. The information required by the $\varepsilon$-machine to track the process is known as the statistical complexity $C_\mu$. This is given by the Shannon entropy (in bits) of the steady-state distribution $P(S)$ over causal states~\cite{crutchfield1989inferring, shalizi2001computational}: 
\begin{equation}
C_\mu=-\sum_{s\in\mathcal{S}}P(s)\log_2(P(s)). 
\end{equation}
This quantity is generically larger than the information shared between the past and future of the process \cite{shalizi2001computational}, indicating that even these optimal models must store redundant information about the past. Indeed, $\varepsilon$-machines are wasteful whenever there is stochasticity in the transition dynamic of the causal states.

\begin{figure}
\begin{tabular}{p{0.328\linewidth} p{0.328\linewidth} p{0.328\linewidth}}
\vspace{0pt} \includegraphics[width=\linewidth]{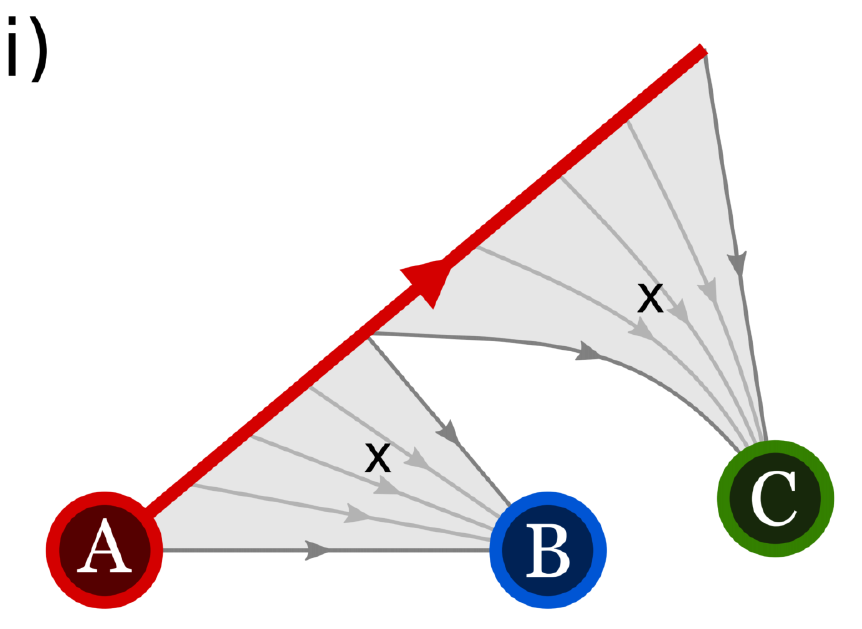}&
\vspace{0pt}\includegraphics[width=\linewidth]{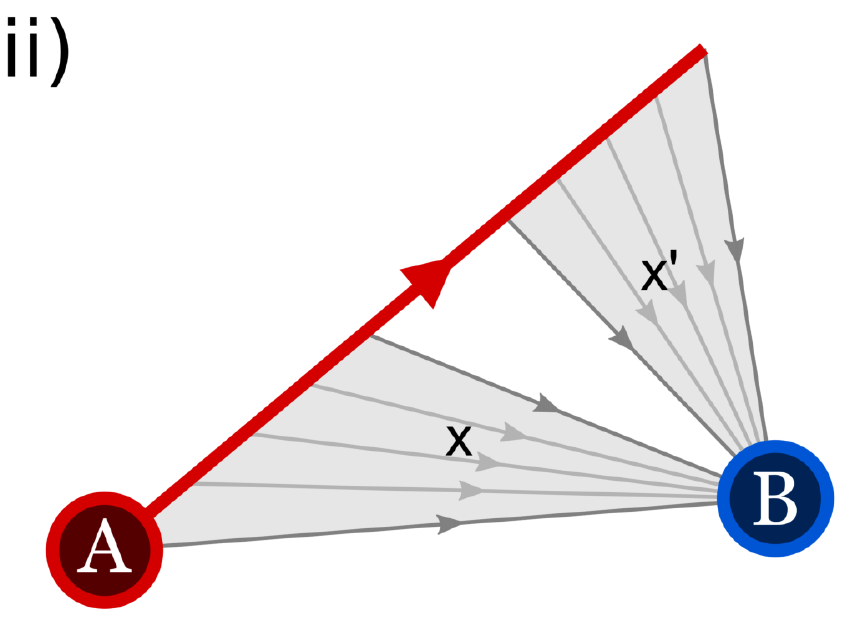}&
\vspace{0pt}\includegraphics[width=\linewidth]{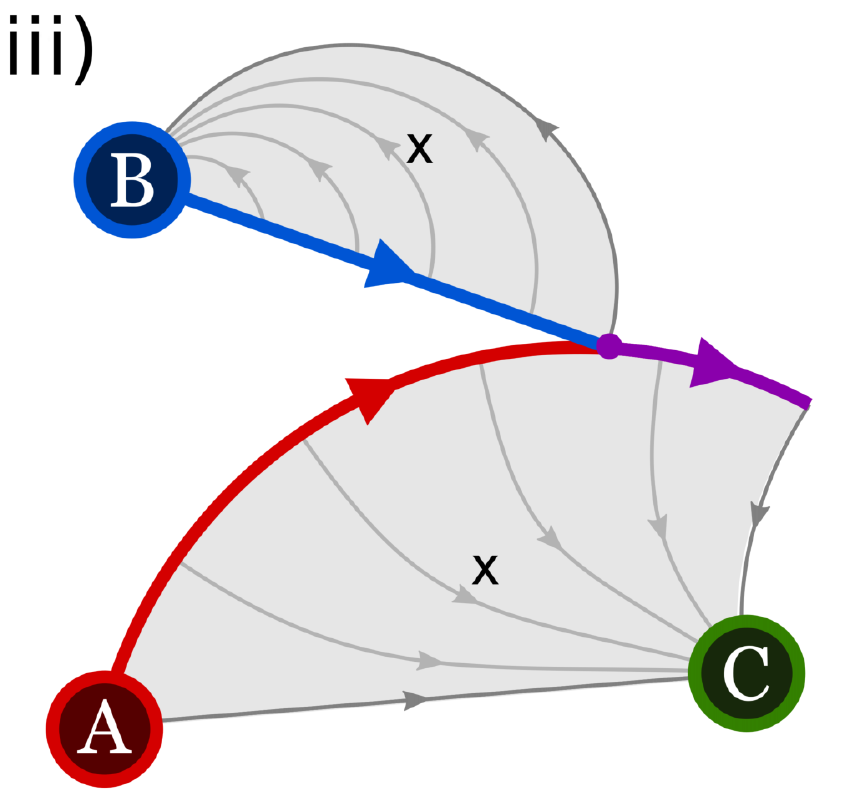}
\end{tabular}
\caption{{\bf Processes without systematic $\varepsilon$-machine construction protocols.} There are currently no systematic construction protocols for the $\varepsilon$-machines of processes where (i) the current mode cannot be synchronised from symbolic dynamics alone; (ii) the modes have symbol-dependent emission-time distributions; or (iii) two modes have identical future dynamics after sufficiently long occupation.}
\label{figrestrictions}
\end{figure}

Though much of the focus of computational mechanics has been on discrete-time symbolic processes, recently analogous results have emerged for continuous-time processes tracking purely temporal~\cite{marzen2017informational}, and both symbolic and temporal dynamics together~\cite{marzen2017structure}. While the optimality proofs hold for such processes, systematic construction protocols for finding causal states are known only for a limited set of processes. The first such class are renewal processes. These describe purely temporal dynamics where all emission symbols are identical, and the times between each consecutive pair of emissions are independent and identically distributed according to a common `waiting time' function~\cite{marzen2017informational}. Recently~\cite{marzen2017structure}, the causal architectures of more general processes with complex symbolic and temporal dynamics have been uncovered, albeit only when they satify certain (quite stringent) restrictions. Assuming the modes $\{g\}$ are already expressed in the minimal unifilar representation, the process must satisfy: 
\begin{quote}
\begin{description}
\item[i)] \hspace{0.45em} Unifilarity (i.e.~synchronisability) of the modes with regards to the observed symbol sequence alone. That is, the mode the system transitions into on the next emission must depend only on the current mode and emitted symbol, and not the time the emission takes place;
\item[ii)] \hspace{0.7em}Emission-time distributions depend only on the current mode [\mbox{$\phi_{kj}^x(t)=\phi_j(t)$}$\forall k,x$]. That is, the time at which emissions occur are independent of the symbol emitted, given the mode;
\item[iii)] \hspace{0.1em} Transition dynamics $T_{kj}^x\phi_j(t)$ must be such that pairs $(g_0,\tpast)$ are not only sufficient for future prediction, but are also minimal. That is, processes where different pairs $(j,\tpast)$ and $(k,\tpast')$ can become causally equivalent are forbidden. This prohibits processes where two modes, conditioned on times since last emission, lead to identical future statistics. This also rules out processes where the conditional emission-time distribution of a mode becomes periodic after a given time since last emission.
\end{description}
\end{quote}
These restrictions are illustrated in \figref{figrestrictions}. They also apply to construction protocols for models tracking both temporal and symbolic dynamics together in discrete-time (to our knowledge, no prior work has explicitly covered this latter scenario).

Our quantum models will not be subject to these restrictions. Moreover, unlike these classical works, we shall not employ differential entropies for continuous-time processes, instead preserving the operational meaning of statistical complexity as the information the model must store about the past. A consequence of this is that whenever the temporal dynamic is not wholly memoryless (i.e.~the $\phi_{kj}^x(t)$ are not all Poisson distributions) $C_\mu$ will diverge, as the $\varepsilon$-machine is ultimately storing a continuous parameter~\cite{marzen2015informational, marzen2017informational, marzen2017structure, garner2017provably, elliott2018superior}.

\section{Quantum models for tracking complex temporal and symbolic dynamics}

While causal states eliminate redundancy in storing information that distinguishes pasts with identical future statistics, they provide no savings when two pasts have similar, yet non-identical futures. This is because states must be either identical or fully distinguishable in classical information theory. In contrast, quantum information~\cite{nielsen2000quantum} can be encoded into states that are only partly distinguishable, and this may be used to reduce the past information that must be retained~\cite{gu2012quantum}. Specifically, causal states can be encoded into quantum states whose overlap increases with the overlap between their corresponding futures. Labelling the information tracked by such quantum models (`q-machines') as $M_q$, we have that generically $M_q\leq C_\mu$, with equality only for processes with no stochasticity in the hidden states of the model~\cite{gu2012quantum} (that is, whenever the $\varepsilon$-machine stores redundant information, a $q$-machine can mitigate some of this redundancy). This reduced entropy bears operational advantages when considering storage or communication of the states of ensembles of simulators of a process (in some cases the advantage may also manifest in the single-shot case~\cite{thompson2018causal}; we do not explore this regime here). The use of $M_q$ rather than $C_q$ denotes that the q-machine may not be optimal, and hence $M_q$ is not necessarily the quantum statistical complexity~\cite{suen2017classical} (but provides an upper bound for it). 

We now provide a systematic protocol that constructs q-machines for simultaneously tracking complex temporal and symbolic dynamics in continuous-time processes. In particular, we show that in general the resultant models:
\begin{quote}
\begin{description}
\item[(C1)] Produce accurate future predictions given the past;
\item[(C2)] Can be operated in a continuous manner;
\item[(C3)] Can be synchronised from the past (are causal);
\item[(C4)] Automatically satisfy causal equivalence relations;
\item[(C5)] Store less information than any classical model.
\end{description}
\end{quote}
As with the classical case, we assume the minimal unifilar modes $\{g\}$ of the process' eeHSMM have been determined. This can be achieved using techniques adapted from discrete-time computational mechanics~\cite{crutchfield1989inferring, shalizi2001causal}, taking the dual $\bm{x}_n$ as effective emitted symbols. Given these minimal modes, a sufficient, causal set of parameters is the pair $(g_0,\tpast)$. A model based on these parameters then need only to identify (with correct probabilities) whether emission occurs in the next infinitesimal interval $dt$, what the emission and subsequent mode are, and to update to be in the corresponding state. The causal states correspond to groupings of such pairs with identical future predictions. Specifically, 
\begin{equation}
(g_0,\tpast)\sim_e(g_0',\tpast')\Leftrightarrow P(G_1,X_0,\Tfut|g_0,\tpast)=P(G_1,X_0,\Tfut|g_0',\tpast'). 
\end{equation}

We introduce several quantities to characterise the processes. First, we define as shorthand for the transition dynamic 
\begin{equation}
\psi_{kj}^x(t)=\sqrt{T_{kj}^x\phi_{kj}^x(t)}.
\end{equation}
Next, the modal steady-state distribution $\pi_j$ is defined as the (unique~\cite{horn1990matrix}) eigenvector of $\sum_xT_{kj}^x$ with unit eigenvalue, normalised such that $\sum_j\pi_j=1$. These $\pi_j$ are the steady-state probabilities that the system is in mode $j$ immediately after an emission. We further define the mode survival probability (the probability that the dwell time in mode $j$ is at least $t$): 
\begin{equation}
\Phi_j(t)=\sum_{xk}\int_t^\infty \psi_{kj}^x(t')^2dt'.
\end{equation}
We also define the mode lifetime (the average dwell time for mode $j$): 
\begin{equation}
\tau_j=\sum_{xk}\int_0^\infty t\psi_{kj}^x(t)^2dt,
\end{equation}
and average emission lifetime $\tau=\sum_j\pi_j\tau_j$. Finally, the modal and mean firing rates are given by reciprocals of the respective lifetimes: $\mu_j=1/\tau_j$ and $\mu=1/\tau$. 

With these definitions, we can express
\begin{equation}
\label{eqcondprob}
P(G_1=k,X_0=x,\Tfut=t'|G_0=j,\Tpast=t)=\frac{\psi_{kj}^x(t+t')^2}{\Phi_j(t)}.
\end{equation}
From these, we define the associated quantum memory states (QMS) for each pair $(g_0,\tpast)=(j,t)$:
\begin{equation}
\ket{\varsigma_j(t)}=\sum_{xk}\int_0^\infty\frac{\psi_{kj}^x(t+t')}{\sqrt{\Phi_j(t)}}dt'\ket{t'}\ket{x}\ket{k},
\end{equation}
where the function inside the integral is the square root of the conditional probability for the future [Eq.~\eqref{eqcondprob}]. The QMS belong to a tripartite composite Hilbert space. The first of these is a continuous space encoding the statistics of the remaining dwell time, while the others are discrete, and correspond to tracking the emitted symbol statistics and subsequent mode respectively. Measurement of the first two spaces in the $\ket{t'}\ket{x}$ basis yields outcomes $t'$ and $x$ with probability density $P(\Tfut=t', X_0=x|G_0=j,\Tpast=t)$, and leaves the final subspace in $\ket{k}$, flagging the mode $k$ to which the system transitions after such an emission. Mapping state $\ket{k}$ (with appropriate blank ancillae) to $\ket{\varsigma_k(0)}$ sets the model in the appropriate post-emission QMS. A measurement sweep of the time subspace over the range $[0,\delta t)$ yields a non-emission result with the correct probability, and for such a non-emission, with a relabelling $t'\to t'-\delta t$, will produce the QMS $\ket{\varsigma_j(t+\delta t)}$. Thus, such measurement sweeps can be used to emulate the passage of time in such models. By performing these measurement sweeps and mappings of flag states for the subsequent post-emission modes, the QMS can model the corresponding stochastic process, fulfilling (C1) and (C2).

As QMS are clearly well-defined for each pair $(g_0,\tpast)$, the q-machine is causal, satisfying (C3). Further, as the QMS depend only on the conditional probabilities that define causal equivalence, they satisfy (C4):
\begin{equation}
(g_0,\tpast)\sim_e(g_0',\tpast')\Leftrightarrow\ket{\varsigma_{g_0}(\tpast)}=\ket{\varsigma_{g_0'}(\tpast')}.
\end{equation}
Thus, QMS corresponding to pasts in the same causal state are identical: \emph{they automatically adopt the causal architecture of the process without the need to explicitly apply the causal equivalence relation}. Further, QMS corresponding to pasts in different causal states generally have non-zero overlap, given by the fidelity of the corresponding conditional probability distributions [Eq.~\eqref{eqcondprob}]. Whenever these overlaps are not all either zero or unity (wherein the $\varepsilon$-machine exhibits no stochasticity in its hidden states, and stores no redundant information~\cite{gu2012quantum}) the QMS steady-state distribution has lower entropy than that of the classical causal states, and \emph{the q-machine stores less information than the corresponding $\varepsilon$-machine}, satisfying (C5).

The information stored in the q-machine may be calculated by determining the Shannon entropy of the spectrum of the steady-state density matrix 
\begin{equation}
\rho_q=\sum_j\int P(j,t)\ket{\varsigma_j(t)}\bra{\varsigma_j(t)}dt,
\end{equation}
where $P(j,t)=\mu\pi_j\Phi_j(t)$ is the steady-state distribution of the QMS (see Appendix A). Using a Gram matrix approach~\cite{horn1990matrix}, we can construct a characteristic equation for the spectrum expressed in terms of the steady-state distribution and overlaps of QMS (see Appendix B):
\begin{equation}
\label{eqcharacteristic}
\mu\sum_{mkx}\sqrt{\pi_j\pi_k}\iint\psi_{mj}^x(t+a)\psi_{mk}^x(t+b) {f_n}_k(b)dtdb=\lambda_n {f_n}_j(a).
\end{equation}
The memory required by the q-machine is then given by 
\begin{equation}
M_q=\sum_n\lambda_n\log_2(\lambda_n).
\end{equation}
These $\lambda_n$ can be determined by solving the integral equation Eq.~\eqref{eqcharacteristic}.

The QMS can be adapted to model coarse-grained versions of a process, wherein continuous time is replaced by small, discrete timesteps. Analogous to classical coarse-graining, in which all states within a given timestep $\delta t$ are merged into a single state prior to applying the causal equivalence relation, we likewise merge the QMS in a particular timestep to form coarse-grained QMS, with probabilities defined by integrals of probability densities over timesteps. The corresponding coarse-grained QMS are given by 
\begin{equation}
\ket{\tilde{\varsigma}_{j}(n)}=\sum_{xk}\sum_{n'}\frac{\tilde{\psi}_{kj}^x(n+n')}{\sqrt{\Phi(n\delta t)}}\ket{n'}\ket{x}\ket{k},
\end{equation}
where $\tilde{\psi}_{kj}^x(n)^2=\int_{n\delta t}^{(n+1)\delta t}\psi_{kj}^x(t)^2dt$. Outcome $n$ from measurement of the first subspace corresponds to $\tfut=n\delta t$. As the coarse-grained states will generally not be mutually orthogonal, a quantum advantage remains. The continuous-time case is recovered in the limit $\delta t\to0$.

For continuously-emitting processes, the emitted symbols are determined immediately after transition events, rather than immediately before. The current emission is then known throughout the current step. An appropriate modification to the protocol is to make a measurement of the symbol subspace in the new QMS immediately after the previous transition event occurs, rather than on the old QMS. The QMS are now also labelled by the current symbol, with 
\begin{equation}
\ket{\varsigma_{jx}(t)}=\sum_k\int\delta_{kj}^x\frac{\sqrt{\phi_{kj}^x(t+t')}}{\sqrt{\Phi_{j}^x(t)}}dt'\ket{t'}\ket{x}\ket{k},
\end{equation}
where $\delta_{kj}^x=1$ if $T_{kj}^x$ is non-zero (and zero otherwise), and $\Phi_{j}^x(t)=\sum_k\int_t^\infty\delta_{kj}^x\phi_{kj}^x(t')dt'$. As with the discrete event case, one can calculate the steady-state density matrix of these states, and from its spectrum determine the amount of information tracked by the q-machine.

\section{Example process with complex temporal and symbolic dynamics exhibiting extreme quantum advantage}

To illustrate our results, we employ our q-machine construction protocol to study an example process that violates the restrictions on current systematic $\varepsilon$-machine construction protocols. Consider the following conceptual scenario allegorising a process: Charlie owns a device that exhibits a constant breaking probability within any fixed interval of time. Upon breakage, he takes the device to either Alice or Bob for repair, after which the device is returned to Charlie. The time taken by Alice and Bob to fix the device is a random variable; Alice's fixing time distribution is different to Bob's, but has the same average. Emissions herald when the device changes hands, and indicate the new holder.

\begin{figure}
\includegraphics[width=\linewidth]{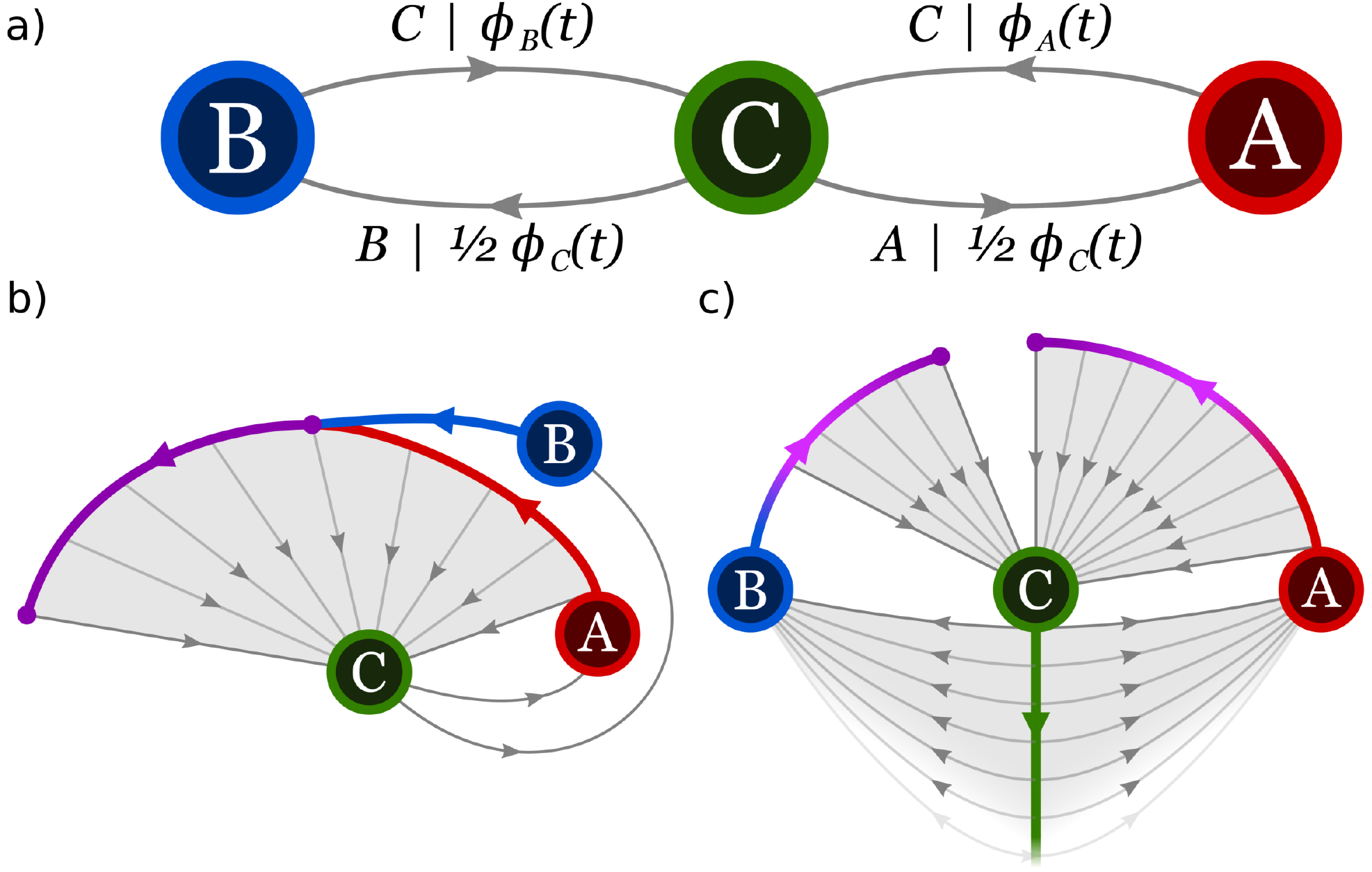}
\caption{{\bf Causal structure of example process.} (a) Example eeHSMM with distributions as given in Eq.~\eqref{eqexampledists}. The temporal dynamics of modes $g_A$ and $g_B$ provide equivalent futures for sufficiently long dwell times, as do all dwell times in mode $g_C$ (see main text). (b) Classically, equivalence relations must be applied manually to appropriately merge the hidden states tracking the dynamics, while (c) quantum models have automatically merged states by construction.}
\label{figexamplecausal}
\end{figure}

Mathematically, this process can be represented by three modes, forming the eeHSMM depicted in \figref{figexamplecausal}(a). Modes $g_A$ and $g_B$ each emit symbol $C$ at times drawn from $\phi_A(t)$ and $\phi_B(t)$ respectively, upon which the system transitions to mode $g_C$. Mode $g_C$ emits symbols $A$ or $B$ with equal probability at times drawn from Poisson distribution $\phi_C(t)$ and transitions into the corresponding mode $g_A$ or $g_B$. We define the distributions:
\begin{align}
\label{eqexampledists}
\phi_A(t)&=\left\{\!\begin{array}{ll} \frac{1}{T_{\mathrm{Fix}}}  & 0\leq t\leq T_{\mathrm{Fix}} \\ 0 & \mathrm{elsewhere}\end{array}\right.;\nonumber\\
\phi_B(t)&=\left\{\begin{array}{ll} \frac{2}{T_{\mathrm{Fix}}}  & \frac{T_{\mathrm{Fix}}}{4}\leq t\leq\frac{3T_{\mathrm{Fix}}}{4} \\ 0 & \mathrm{elsewhere}\end{array}\right.;\nonumber\\
\phi_C(t)&=\frac{e^{-t/T_{\mathrm{Brk}}}}{T_{\mathrm{Brk}}}.
\end{align} 

We see that this process exhibits causal equivalence between certain of its causal pairs, and as such falls outside the class for which systematic $\varepsilon$-machine construction methods are currently known. Specifically, because $\phi_C(t)$ is a Poisson distribution, all dwell times within this mode have identical conditional futures, and so only a single causal state is needed to describe occupation of this mode~\cite{marzen2015informational, marzen2017informational, elliott2018superior}:
\begin{equation}
(g_C,t)\sim_e(g_C,t')\forall t,t'\geq0.
\end{equation}
Further, we see that the conditional futures for modes $g_A$ and $g_B$ can become identical for certain combinations of dwell times; specifically, we have 
\begin{equation}
(g_A,t+\frac{T_{\mathrm{Fix}}}{2})\sim_e(g_B,t+\frac{T_{\mathrm{Fix}}}{4})\forall t\geq0.
\end{equation}
These both exemplify violations of restriction iii. The conceptual scenario can be easily extended to describe a process in violation of restriction ii: suppose Charlie's device has two possible faults, with different failure rates, and the choice between Alice and Bob is decided by which of the faults occured. Restriction i would be violated if Charlie were to merely announce that the device has broken (instead of who will be fixing it) and the choice between Alice and Bob is determined by how long the device took to fail.

The relative simplicity of this example allows us to perform the appropriate application of the equivalence relations to merge causal pairs into causal states on an \emph{ad hoc} basis, as is displayed in \figref{figexamplecausal}(b). Notably however, in contrast to this we can exploit the self-merging nature of the QMS construction, and blindly assign QMS for each causal pair. These QMS will automatically satisfy the causal equivalence relations, and will not incur any penalty in the information the q-machine must track. This is depicted in \figref{figexamplecausal}(c).

\begin{figure}
\includegraphics[width=0.495\linewidth]{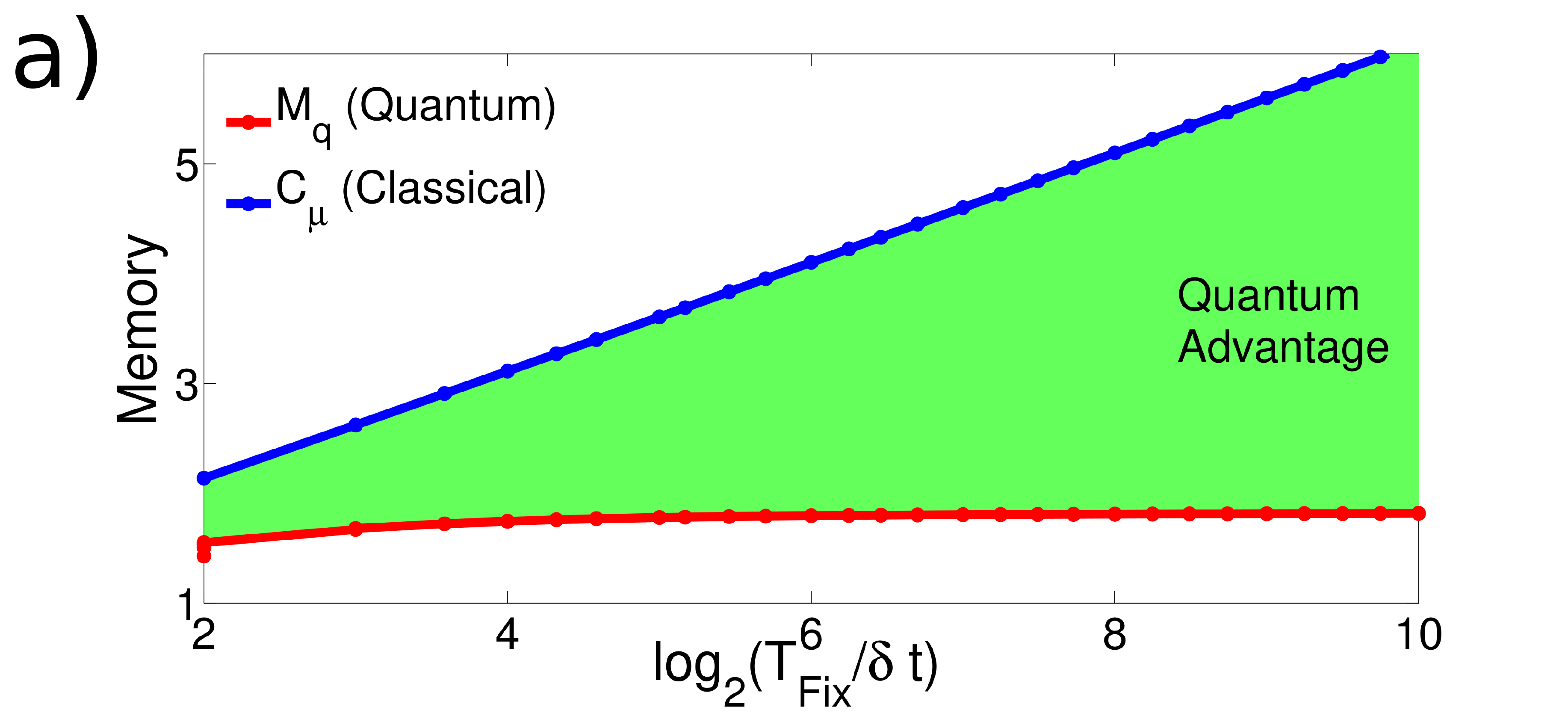}
\includegraphics[width=0.495\linewidth]{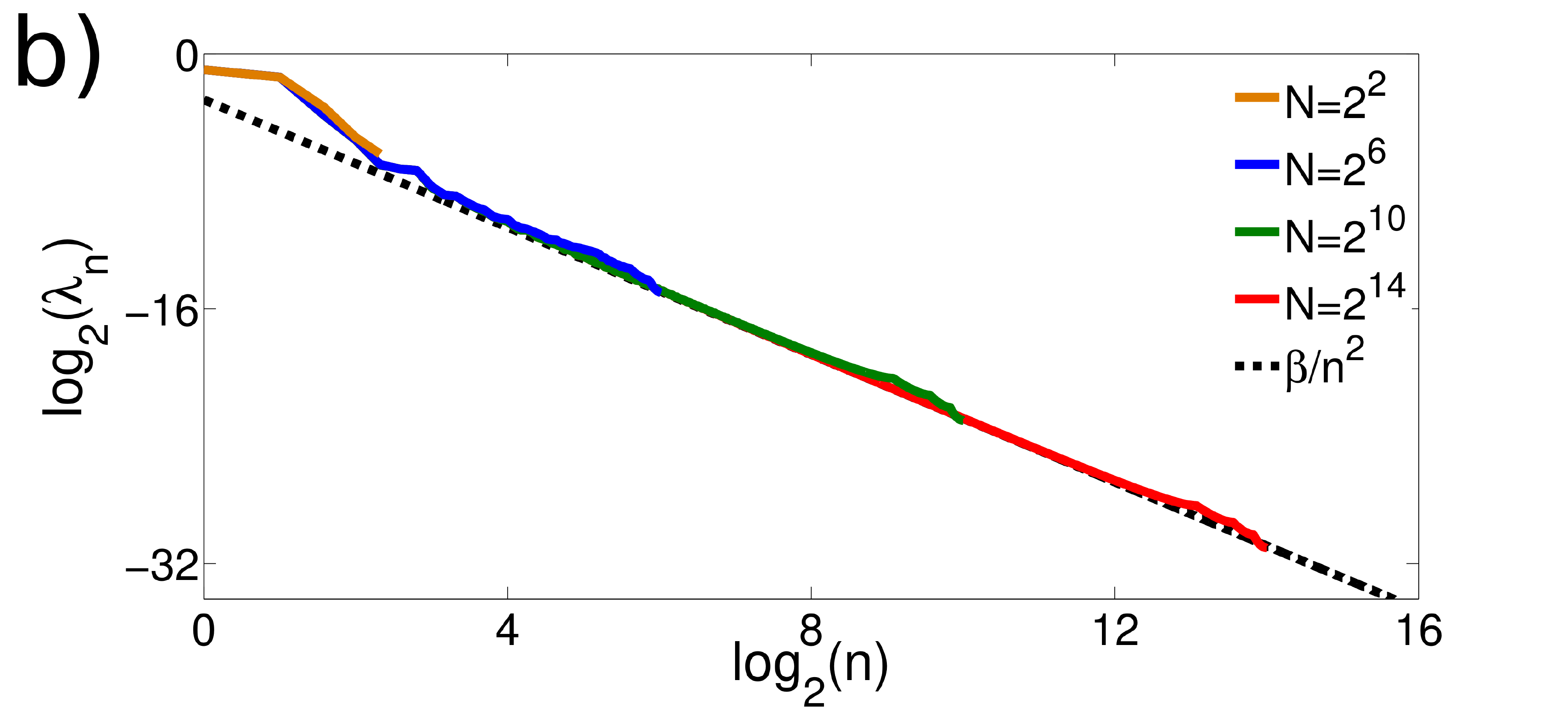}
\caption{
{\bf Memory requirements for example process.}  (a) The information stored within the $\varepsilon$-machine and q-machine at increasing levels of precision, showing a quantum advantage that appears to become unbounded in the continuum limit (plot shown for $\tau=2T$, with data points taken at values $\tau/4\delta t\in\mathbb{Z}^+$ for calculational simplicity). (b) Inspection of the steady-state spectrum for increasingly fine discretisation ($N=T_{\mathrm{Fix}}/\delta t$) of the q-machine indicates that the eigenvalues appear to fall off with a $1/n^2$ dependence. Here $\beta$ is a normalisation constant chosen such that $\sum_{n=1001}^{\infty}\beta/n^2=\sum_{n=1001}^{N+1}\lambda_n$ for the $N=2^{14}$ case (eigenvalues ranked largest to smallest).}
\label{figexamplememory}
\end{figure}

As described in Appendix C, we can calculate the relevant modal and mean lifetimes and firing rates of the process, the coarse-grained causal states and QMS, and their respective steady-state probabilities. Solving the course-grained analogue of Eq.~\eqref{eqcharacteristic} we can determine the steady-state density matrix spectrum, and hence calculate $M_q$ for the particular level of coarse-graining. We can further calculate $C_\mu$ at the same coarse-graining for comparison. In \figref{figexamplememory}(a) we show that with increasingly fine coarse-graining $C_\mu$ diverges, while $M_q$ appears to converge towards a finite value. This suggests an unbounded advantage of the q-machine over the $\varepsilon$-machine, similar to that seen for renewal processes~\cite{elliott2018superior}. Inspection of the steady-state density matrix spectrum indicates that it falls off approximately as $1/n^2$ [\figref{figexamplememory}(b)]. When the spectrum has this dependence, the associated entropy is finite~\cite{elliott2018superior} providing further support that $M_q$ is bounded in the continuum limit.

\section{Discussion}

We have proposed a systematic construction protocol for quantum models of general complex continuous-time stochastic processes that automatically adopt the process' causal architecture, and exhibit an entropic memory advantage over their optimal classical counterparts. Moreover, our model construction protocol can be applied more generally than corresponding protocols for optimal classical models, without any restrictions on the combined symbolic and temporal dynamics. This allows our models to be utilised to study a much broader range of processes, such as the aforementioned examples in our introduction~\cite{yu2010hidden, levinson1986continuously, rabiner1989tutorial, bulla2006stylized, gerstner2002spiking, marzen2015time, kulp1996generalized, yu2002hidden, garavaglia2011earthquake}.

While the information cost of of a quantum model constructed by our protocol is lower than any classical model, there is no claim of optimality over all quantum models. By accounting for longer-range temporal correlations~\cite{mahoney2016occam, riechers2016minimized, binder2018practical}, it has been found that discrete-time q-machines can reduce their information cost, and analogous constructs may be possible in the continuous-time case, providing further memory savings. Other reductions may be possible by exploiting the possibility of using complex amplitudes in the QMS. Even so, our current models provide an upper bound on the information cost for the optimal quantum model, and a lower bound is given by the mutual information between the past and future of the process~\cite{shalizi2001computational,gu2012quantum}.

It is interesting to consider how to implement the q-machine as a simulator. Our earlier work on q-machines for continuous-time renewal processes~\cite{elliott2018superior} discussed how they might be realised by using the position of a particle as the continuous variable tracking time, and the motion of the particle towards a detector as the measurement sweep. One can envisage the more general multi-mode, multi-symbol processes here might be implemented similarly, using internal states of the particle for the mode and symbol subspaces of the QMS. We leave specific details of particular experimental implementations as an open question for future work.

Finally, let us remark on the apparent unbounded advantage in our example. In earlier work we have postulated that this is likely a typical feature of continuous-time q-machines~\cite{elliott2018superior}. We recapitulate the argument~\cite{garner2017provably, elliott2018superior} here. Consider a coarse-graining with timesteps $\delta t$. With further refinement, the new QMS will typically have large overlap with the existing temporally-adjacent QMS. At very fine coarse-graining, the new states will be almost identical to those existing, and so the increase in information cost will be ever-decreasing, resulting in the observed convergence of $M_q$. In contrast, the mutual orthogonality of the classical states leads to a logarithmic divergence in information cost. An enticing problem for future work is to develop methods of coarse-graining that exploit such quantum features. We expect that such a quantum coarse-graining could be used for near-exact simulation with extreme memory advantages even in the single-shot regime. It would also be interesting to consider the extension to input-output processes~\cite{barnett2015computational, thompson2017using} operating in continuous-time.

\section*{Acknowledgements}
This work was funded by the Lee Kuan Yew Endowment Fund (Postdoctoral Fellowship), Singapore Ministry of Education Tier 1 grant RG190/17, FQXi Large Grants: ``The role of quantum effects in simplifying adaptive agents" and ``Observer-dependent complexity: The quantum-classical divergence over `what is complex?'", John Templeton Foundation grant 53914, and Singapore National Research Foundation Fellowship NRF-NRFF2016-02. T.J.E.~thanks the Centre for Quantum Technologies for their hospitality.

\appendix

\section{Derivation of steady-state distribution for causal pairs}

We here derive the steady-state probability distribution of the causal pair $(j,t)$, where $j$ is the current mode, and $t$ the time since last emission. As stated in the main text, this distribution is given by 
\begin{equation}
P(G_0=j,\Tpast=t)=\mu\pi_j\Phi_j(t).
\end{equation}
Let us first decompose 
\begin{equation}
P(G_0=j,\Tpast=t)=P(\Tpast=t|G_0=j)P(G_0=j),
\end{equation}
and find each of the constituent probabilities. 

Beginning with the former, recall that the conditional probability density of emission from mode $j$ at time $t'$ from the present given a prior wait time of $t$ is given by 
\begin{equation}
P(\Tfut=t'|\Tpast=t,G_0=j)=\sum_{kx}\frac{T_{kj}^x\phi_{kj}^x(t+t')}{\Phi_j(t)}.
\end{equation}
The conditional survival probability in mode $j$ to time $t+t'$ is thus given by $\Phi_j(t+t')/\Phi_j(t)$. Clearly then, the probability of progressing without an emission attenuates by the survival probability, and we have 
\begin{equation}
P(\Tpast=t+t'|G_0=j)=\frac{\Phi_j(t+t')}{\Phi_j(t)}P(\Tpast=t|G_0=j).
\end{equation}
This is satisfied for all $t,t'\geq0$, and hence 
\begin{equation}
P(\Tpast=t|G_0=j)=\Phi_j(t)P(\Tpast=0|G_0=j)\quad\forall t\geq0.
\end{equation}
We can further find $P(\Tpast=0|G_0=j)$ by normalising this distribution:
\begin{equation}
P(\Tpast=0|G_0=j)^{-1}=\int_0^\infty\Phi_j(t)dt=\mu_j^{-1}.
\end{equation}
Thus, 
\begin{equation}
\label{eqsteadytime}
P(\Tpast=t|G_0=j)=\mu_j\Phi_j(t).
\end{equation}

Next, we must determine the probability of a particular mode being occupied in the steady-state. Recall that $\pi_j$ is the steady-state probability that after an emission the system transitions into mode $j$. Since the expected time of remaining in mode $j$ is given by the modal lifetime $\tau_j$, we have that the steady-state probability of being in mode $j$ is proportional to $\pi_j\tau_j$. By noticing that the sum of this over all modes is the average emission lifetime $\tau$, we hence obtain 
\begin{equation}
\label{eqsteadymode}
P(G_0=j)=\frac{\pi_j\tau_j}{\tau}=\frac{\mu\pi_j}{\mu_j}.
\end{equation}

Thus, putting Eqs.~\eqref{eqsteadytime} and \eqref{eqsteadymode} together, we have
\begin{equation}
P(G_0=j,\Tpast=t)=\mu\pi_j\Phi_j(t).
\end{equation}

\section{Gram matrix and derivation of the characteristic equation}

A useful tool for calculating the spectrum of a density matrix is offered by the Gram matrix~\cite{horn1990matrix}. The Gram matrix provides an alternative operator that possesses the same spectrum as the original operator of interest, but that may be more amenable for extracting said spectrum. To obtain the Gram matrix, we consider a purification of the QMS steady-state ensemble: 
\begin{equation}
\ket{\Psi}_{qG}=\sum_j\int\sqrt{P(j,t)}\ket{\varsigma_j(t)}\ket{t}\ket{j}dt.
\end{equation}
By performing a partial trace over the second and third subsystems we obtain the original steady-state density matrix $\rho_q$. On the other hand, tracing out the first subsystem leaves us with
\begin{equation}
\rho_G=\sum_{jj'}\iint\sqrt{P(j,t)P(j',t')}\braket{\varsigma_j(t)}{\varsigma_{j'}(t')}\ket{t}\ket{j}\bra{t'}\bra{j'}dtdt'.
\end{equation}
The resulting operator $\rho_G$ is called the Gram matrix, and as noted above, has the same spectrum as $\rho_q$. It has a corresponding characteristic equation 
\begin{equation}
\rho_Gf_n=\lambda_n f_n,
\end{equation}
where $f_n$ are the eigenfunctions, and $\lambda_n$ the spectrum. Expanding this in terms of the QMS and steady-state distributions, we obtain the characteristic equation Eq.~\eqref{eqcharacteristic}, which may be solved to obtain the spectrum $\lambda_n$, and hence calculate the information $M_q$ stored by the q-machine.

\section{Further calculational details for the example process}
\begin{figure*}
\includegraphics[width=0.32\linewidth]{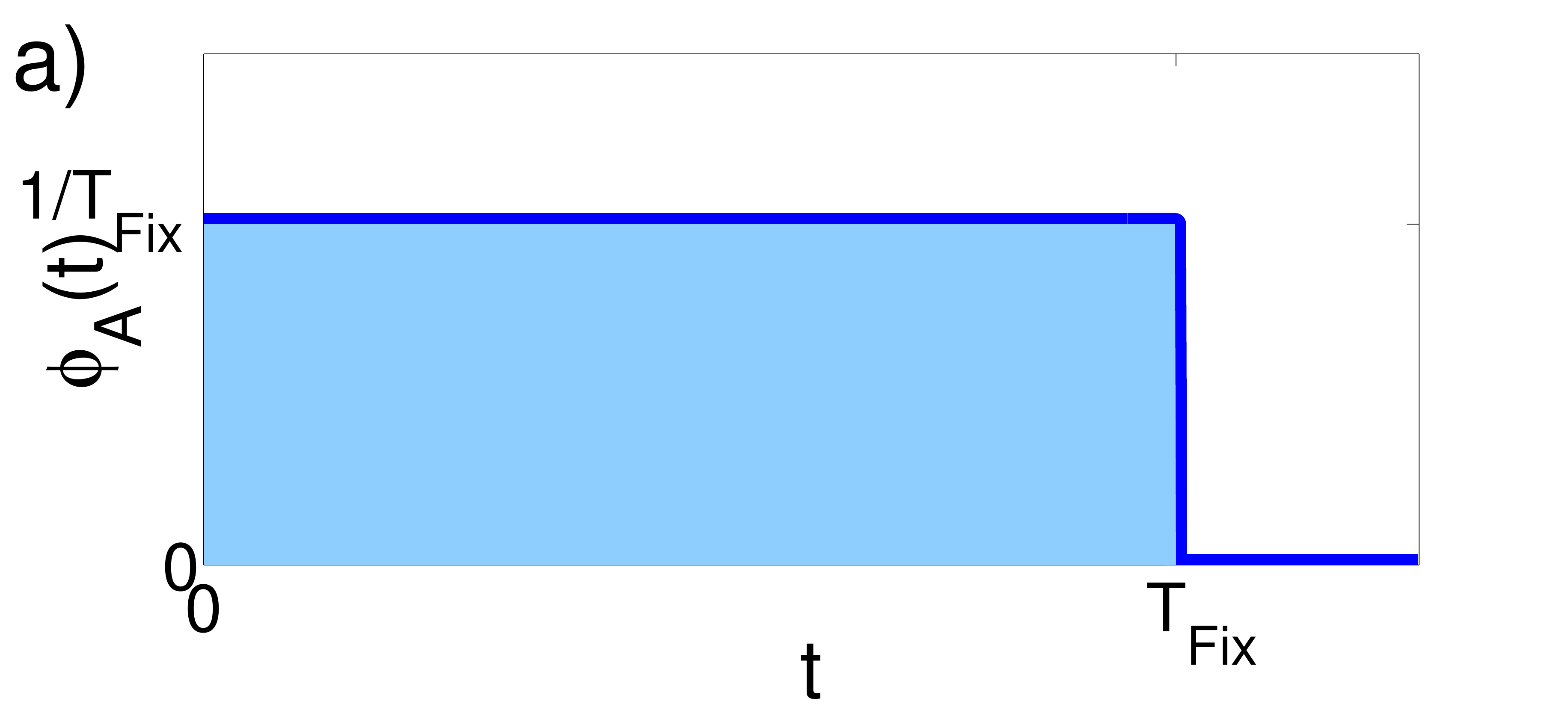}
\includegraphics[width=0.32\linewidth]{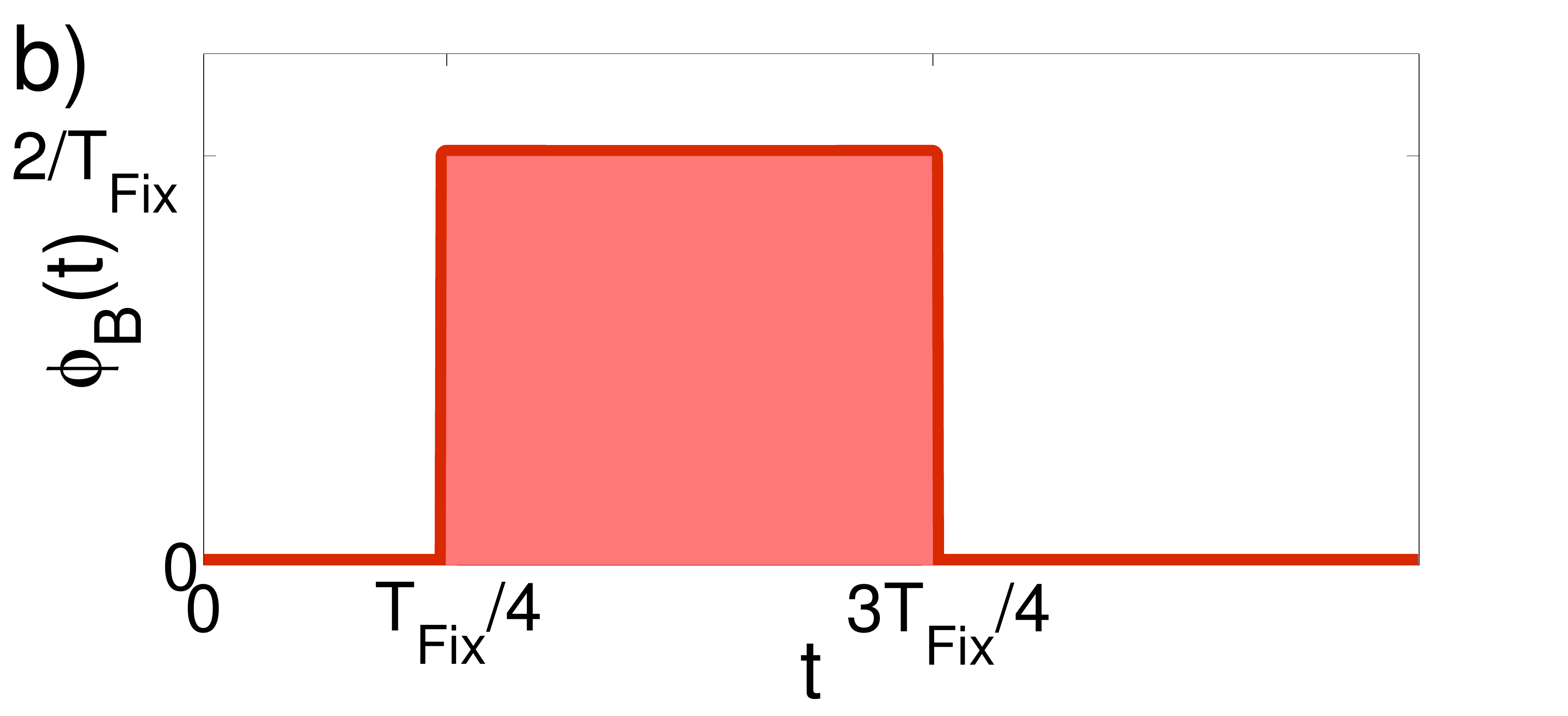}
\includegraphics[width=0.32\linewidth]{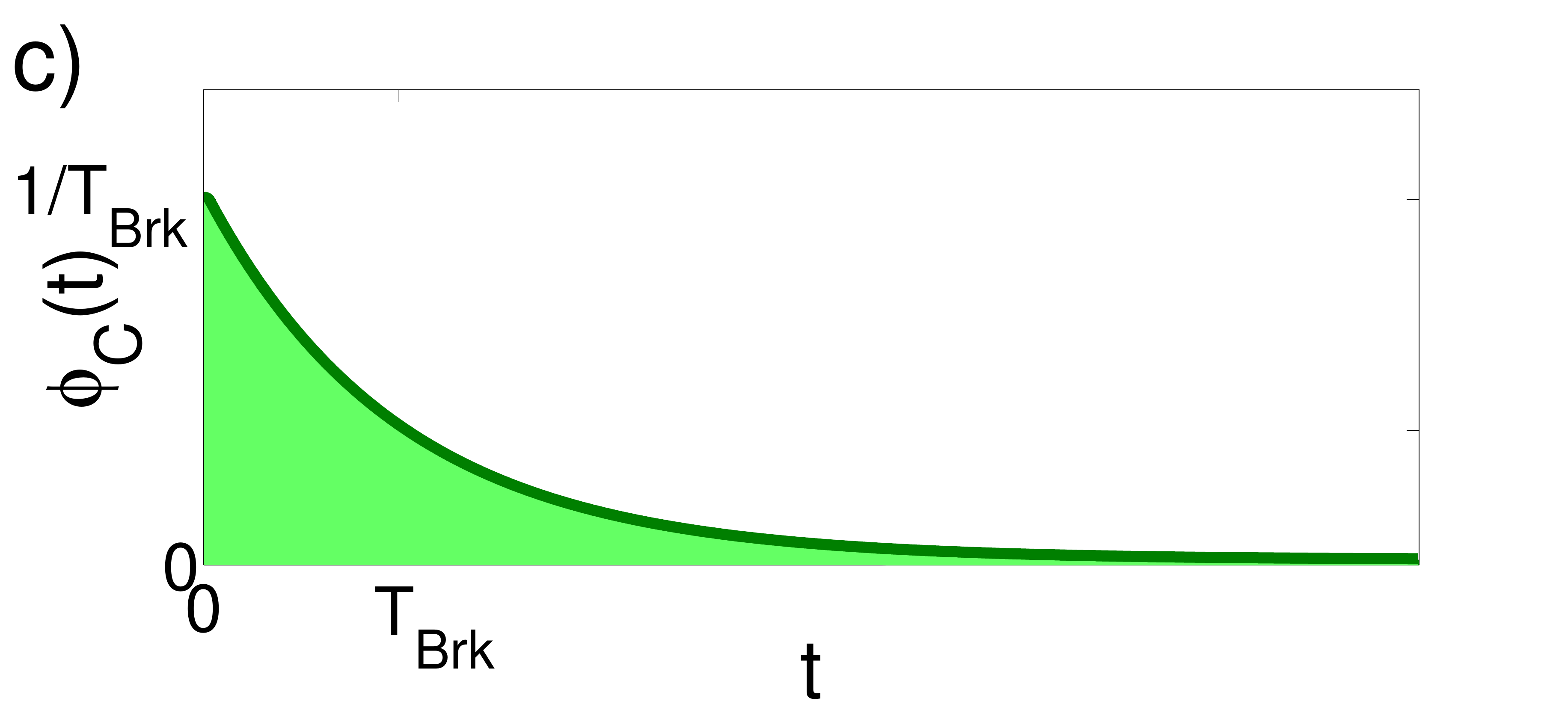}
\caption{{\bf Emission distributions for example.} Dwell times of Eqs.~\eqref{eqexampledistssupp} describing the temporal dynamics of the example.}
\label{figexampledists}
\end{figure*}

Here we provide the detailed derivations of the properties of the example in the main text. Recall the definition of the process, as depicted in \figref{figexamplecausal}(a), whereby two modes $g_A$ and $g_B$ each emit symbol $C$ while transitioning to mode $g_C$ at a time drawn from $\phi_A(t)$ or $\phi_B(t)$ respectively, while mode $g_C$ emits $A$ or $B$ and transitions to the corresponding mode with equal probability, at a time drawn from $\phi_C(t)$. The respective distributions are defined by
\begin{align}
\label{eqexampledistssupp}
\phi_A(t)&=\left\{\begin{array}{ll} \frac{1}{T_{\mathrm{Fix}}}  & 0\leq t\leq T_{\mathrm{Fix}} \\ 0 & \mathrm{elsewhere}\end{array}\right.;\nonumber\\
\phi_B(t)&=\left\{\begin{array}{ll} \frac{2}{T_{\mathrm{Fix}}}  & \frac{T_{\mathrm{Fix}}}{4}\leq t\leq\frac{3T_{\mathrm{Fix}}}{4} \\ 0 & \mathrm{elsewhere}\end{array}\right.;\nonumber\\
\phi_C(t)&=\frac{e^{-t/T_{\mathrm{Brk}}}}{T_{\mathrm{Brk}}},
\end{align}
as illustrated in \figref{figexampledists}.

From these distributions, we can straightforwardly calculate the modal lifetimes and steady-state distributions using the definitions in the main text. These are as follows: 
\begin{align}
\tau_A=\tau_B=\frac{T_{\mathrm{Fix}}}{2};\qquad&
\tau_C=T_{\mathrm{Brk}};\nonumber\\
\pi_A=\pi_B=\frac{1}{4};\qquad&
\pi_C=\frac{1}{2}.
\end{align}
The average lifetime is $\tau=(T_{\mathrm{Fix}}+2T_{\mathrm{Brk}})/4$. We can also calculate the mode survival probabilities:
\begin{align}
\Phi_A(t)&=\left\{\begin{array}{ll} 1-\frac{t}{T_{\mathrm{Fix}}}  & 0\leq t\leq T_{\mathrm{Fix}} \\ 0 & \mathrm{elsewhere}\end{array}\right.;\nonumber\\
\Phi_B(t)&=\left\{\begin{array}{lll} 1 & 0\leq t <\frac{T_{\mathrm{Fix}}}{4} \\ \frac{3}{2} - \frac{2t}{T_{\mathrm{Fix}}}  & \frac{T_{\mathrm{Fix}}}{4}\leq t\leq\frac{3T_{\mathrm{Fix}}}{4} \\ 0 & \mathrm{elsewhere}\end{array}\right.;\nonumber\\
\Phi_C(t)&=e^{-t/T_{\mathrm{Brk}}}.
\end{align}

The corresponding steady-state probabilities are:
\begin{align}
\label{eqsteadystateexprobs}
P(g_A,t)&=\left\{\begin{array}{ll} \dfrac{1-\frac{t}{T_{\mathrm{Fix}}}}{T_{\mathrm{Fix}}+2T_{\mathrm{Brk}}}  & 0\leq t\leq T_{\mathrm{Fix}} \\ 0 & \mathrm{elsewhere}\end{array}\right.;\nonumber\\
P(g_B,t)&=\left\{\begin{array}{lll} \dfrac{1}{T_{\mathrm{Fix}}+2T_{\mathrm{Brk}}} & 0\leq t <\frac{T_{\mathrm{Fix}}}{4} \vspace{0.5em}\\ \dfrac{\frac{3}{2} - \frac{2t}{T_{\mathrm{Fix}}}}{T_{\mathrm{Fix}}+2T_{\mathrm{Brk}}}  & \frac{T_{\mathrm{Fix}}}{4}\leq t\leq\frac{3T_{\mathrm{Fix}}}{4} \vspace{0.5em}\\ 0 & \mathrm{elsewhere}\end{array}\right.;\nonumber\\
P(g_C,t)&=\frac{2e^{-t/T_{\mathrm{Brk}}}}{T_{\mathrm{Fix}}+2T_{\mathrm{Brk}}}.
\end{align}

While no general prescription is presently known for implementing the causal equivalence relations for continuous-time processes in classical systems~\cite{marzen2017structure}, their relative simplicity in this example allows for them to be applied on an \emph{ad hoc} basis, as displayed in \figref{figexamplecausal}(b). We can merge all times within mode $g_C$ into a single causal state, as well as all pairs of states $(g_A,t+T_{\mathrm{Fix}}/2)$ and $(g_B,t+T_{\mathrm{Fix}}/4)$ for $t\geq0$. For clarity, in the following we label states by a specific mode only for the unmerged states, and use $g_M$ to represent states in the merge between the $g_A$ and $g_B$ pairs. The corresponding causal states have steady-state probabilities given by
\begin{align}
P(A,t)&=\frac{1-\frac{t}{T_{\mathrm{Fix}}}}{T_{\mathrm{Fix}}+2T_{\mathrm{Brk}}} \quad 0\leq t<\frac{T_{\mathrm{Fix}}}{2} \nonumber\\
P(B,t)&=\frac{1}{T_{\mathrm{Fix}}+2T_{\mathrm{Brk}}} \quad 0\leq t<\frac{T_{\mathrm{Fix}}}{4} \nonumber\\
P(M,t)&=\frac{3\left(\frac{1}{2}-\frac{t}{T_{\mathrm{Fix}}}\right)}{T_{\mathrm{Fix}}+2T_{\mathrm{Brk}}} \quad 0\leq t<\frac{T_{\mathrm{Fix}}}{2} \nonumber\\
P(C)&=\frac{2T_{\mathrm{Brk}}}{T_{\mathrm{Fix}}+2T_{\mathrm{Brk}}}.
\end{align}

As discussed in the main text, the continuous nature of the distributions leads to the classical statistical complexity $C_\mu$ being infinite. However, we can still investigate how the complexity grows with increasingly fine discretisation, and compare this to the quantum case. For calculational simplicity, we use timesteps $\delta t$ such that $T_{\mathrm{Fix}}/4\delta t$ is an integer, which ensures that each timestep does not span across any borders where the behaviour of the probability distributions change (i.e.~there are no `partially merged' timesteps). When discretising, we note that there is a small $O(\delta t)$ correction to the effective lifetimes and firing rates~\cite{marzen2015informational}, which can be appropriately accounted for by inclusion of a normalisation factor $\mathcal{N}$ such that $\mathcal{N}\sum_{n=0}^\infty\Phi(n\delta t)=\mu\pi_j\tau_j$. Calculating the Shannon entropy of the discretised distributions, we obtain $C_\mu$ at each level of precision, as plotted in \figref{figexamplememory}(a) for the case $T_{\mathrm{Fix}}=2T_{\mathrm{Brk}}$.

For the q-machine, we must first construct the discretised QMS, as prescribed in the main text:

\begin{widetext}
\begin{equation}
\begin{array}{ll}
\ket{\tilde{\varsigma}_{g_A}(n)}=\sum_{n'=0}^{\frac{T_{\mathrm{Fix}}}{\delta t}-(n+1)}\dfrac{\sqrt{\delta t}}{\sqrt{\frac{T_{\mathrm{Fix}}}{\delta t}-n}}\ket{n'}\ket{C}\ket{g_C} & n=0,1\ldots,\frac{T_{\mathrm{Fix}}}{\delta t}-1 \\
\ket{\tilde{\varsigma}_{g_B}(n)}=\sum_{n'=\mathrm{max}(0,\frac{T_{\mathrm{Fix}}}{4\delta t}-n)}^{\frac{3T_{\mathrm{Fix}}}{4\delta t}-(n+1)}\dfrac{\sqrt{\delta t}}{\sqrt{\frac{3T_{\mathrm{Fix}}}{4\delta t}-\mathrm{max}(\frac{T_{\mathrm{Fix}}}{4\delta t},n)}}\ket{n'}\ket{C}\ket{g_C}& n=0,1,\ldots,\frac{3T_{\mathrm{Fix}}}{4\delta t} \\
\ket{\tilde{\varsigma}_{g_C}(n)}=\sum_{n'=0}^\infty\dfrac{e^{-n'\delta t/2T_{\mathrm{Brk}}}\sqrt{1-e^{-\delta t/T_{\mathrm{Brk}}}}}{\sqrt{2}}\ket{n'}(\ket{A}\ket{g_A}+\ket{B}\ket{g_B})& n\in\mathbb{N}.
\end{array}
\end{equation}
\end{widetext}

Due to the self-merging nature of the QMS, this blind construction protocol is automatically consistent with the causal equivalence relations satisfied by the classical causal states. By constructing a density matrix of these QMS, with weighting given by the appropriate steady-state probabilities Eqs.~\eqref{eqsteadystateexprobs} (or rather, discretised analogues thereof), we can find its spectrum, and the associated Shannon entropy provides the quantum memory requirement $M_q$. These are plotted for increasingly fine levels of discretisation in \figref{figexamplememory}(a) alongside the corresponding classical memory requirements, where it is clear that the q-machine requires less information than the optimal classical machine, and appears to converge to a bounded value, much like was previously found for renewal processes~\cite{elliott2018superior}. 

\bibliography{ref}

\end{document}